\documentclass[prb,twocolumn,groupedaddress, showkeys,showpacs]{revtex4}

\usepackage{graphicx}
\usepackage{amsmath}
\usepackage{mathrsfs}

\begin{document}


\title{Transport of active ellipsoidal particles in ratchet potentials}


\author{Bao-quan  Ai} \email[Email: ]{aibq@scnu.edu.cn}
\author{Jian-chun Wu}


\affiliation{Laboratory of Quantum Engineering and Quantum Materials, School of Physics and Telecommunication
Engineering, South China Normal University, 510006 Guangzhou, China.}


\date{\today}
\begin{abstract}
  \indent Rectified transport of active ellipsoidal particles is numerically investigated in a two-dimensional asymmetric potential. The out-of-equilibrium condition for the active particle is an intrinsic property, which can break thermodynamical equilibrium and induce the directed transport. It is found that the perfect sphere particle can facilitate the rectification, while the needlelike particle destroys the directed transport. There exist optimized values of the parameters (the self-propelled velocity, the torque acting on the body) at which the average velocity takes its maximal value.  For the ellipsoidal particle with not large asymmetric parameter, the average velocity decreases with increasing the rotational diffusion rate, while for the needlelike particle (very large asymmetric parameter), the average velocity is a peaked function of the rotational diffusion rate. By introducing a finite load, particles with different shapes (or different self-propelled velocities) will move to the opposite directions, which is able to separate particles of different shapes (or different self-propelled velocities).
\end{abstract}

\pacs{05. 60. Cd, 05. 40. -a, 82. 70. Dd}



\maketitle
\section {Introduction}
\indent Noise-induced transport far from equilibrium plays a crucial role in many processes from physical and biological to social systems. The transport properties of systems consisting of active particles have generated much attention. There are numerous realizations of active particles\cite{lauga,toner} in nature ranging from bacteria \cite{leptos,shenoy,hill,diluzio} and spermatozoa\cite{riedel} to artificial colloidal microswimmers. Self-propulsion is an essential feature of most living systems, which can maintain metabolism and perform movement. The kinetic of self-propulsion particles moving in potentials could exhibit peculiar behavior \cite{Burada,Schweitzer,tailleur,Schimansky-Geier,kaiser,fily,buttinoni, bickel,mishra,czirok, peruani,weber,stark,ai1,chen}.
The problem of rectifying motion in random environments is an important issue, which has many theoretical and practical implications\cite{rmp}. At equilibrium the periodic potential alone is not able to produce a rectification effect, due to the detailed balance preventing time symmetry breaking. Indeed, one has to add some perturbation which breaks the time symmetry and brings the system out of equilibrium. For the active particles, the out-of-equilibrium condition is an intrinsic property of the system. So the active particle without any external forces can break the symmetry of the system and be rectified in periodic systems.

\indent Recently, rectification of self-propelled particles in
asymmetric external potentials has attracted much attention.
Angelani and co-workers \cite{angelani} studied the run-and tumble
particles in periodic potentials and found that the asymmetric
potential produces a net drift speed.  Even in the symmetric
potential a spatially modulated self-propulsion and a phase shift
against the potential can induce the directed transport
\cite{potosky}. Recently, transport of Janus particles in
periodically compartmentalized channel is investigated \cite{ghosh}
and the rectification can be orders of magnitude stronger than that
for ordinary thermal potential ratchets. In all these
studies\cite{angelani,potosky,ghosh} on active ratchet, the active
particle was treated as the point spherical particle.

\indent However, shape deformation of particles plays an important role in nonequilibrium transport processes \cite{Grima,Han,Ohta,Matsuo,Mammadov,Gralinski}. Han and co-workers \cite{Han} experimentally studied the Brownian motion of isolated ellipsoid particles in two dimensions and quantified the crossover from short-time anisotropic to long-time isotropic diffusion.  In the presence of an external potential, the external force can amplify the non-Gaussian character of the spatial probability distributions \cite{Grima}. Ohta and co-workers \cite{Ohta} found that an isolated deformable particle exhibits a bifurcation such that a straight motion becomes unstable and a circular motion appears. Due to the coupling of the rotational and translational motion, Brownian motion of asymmetrical particles is considerably more complicated compared to the spherical case, and thus shows peculiar behavior. Therefore, how active asymmetrical particles are rectified from a ratchet potential may receive much attention. In this paper, we will extend the study of active ratchet from the spherical particle to the asymmetric particle. We focus on finding how the asymmetry of the particle affects the rectified transport and how asymmetric particles can be separated.

\section{Model and methods}
\indent We consider an ellipsoidal (anisotropic) particle moving in
a two-dimensional ratchet potential (shown in Fig. 1). The particle
is self-propelled along its long axis. In the lab $x$-$y$ frame, the
particle at a given time $t$ can be described by the position vector
$\vec{R}(t)$ of its center of mass, which also corresponds to the
coordinates in the body frame ($\delta\hat{x}$, $\delta\hat{y}$).
$\theta(t)$ is the angle between the $x$ axis of the lab frame and
the $\hat{x}$ axis of the body frame. Rotational and translational
motion in the body frame are always decoupled, so the dynamics of
the active ellipsoid particle is described by the Langevin equations
in this frame \cite{Grima,Han},
\begin{figure}
  \centering
  \vspace{-0.5cm}
  \hspace{-1cm}
  \includegraphics[width=10cm]{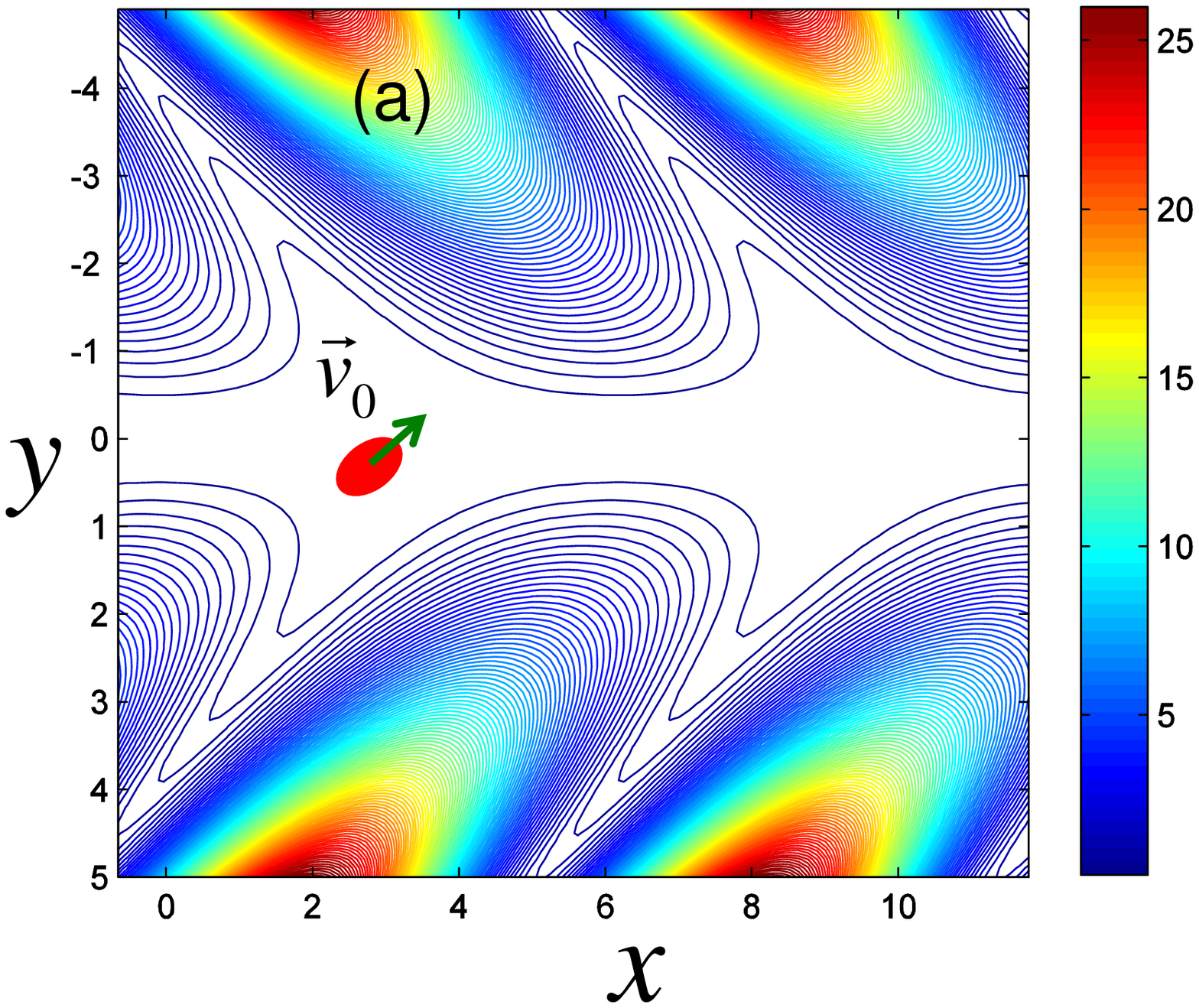}
   \vspace{-0.5cm}
  \includegraphics[width=10cm]{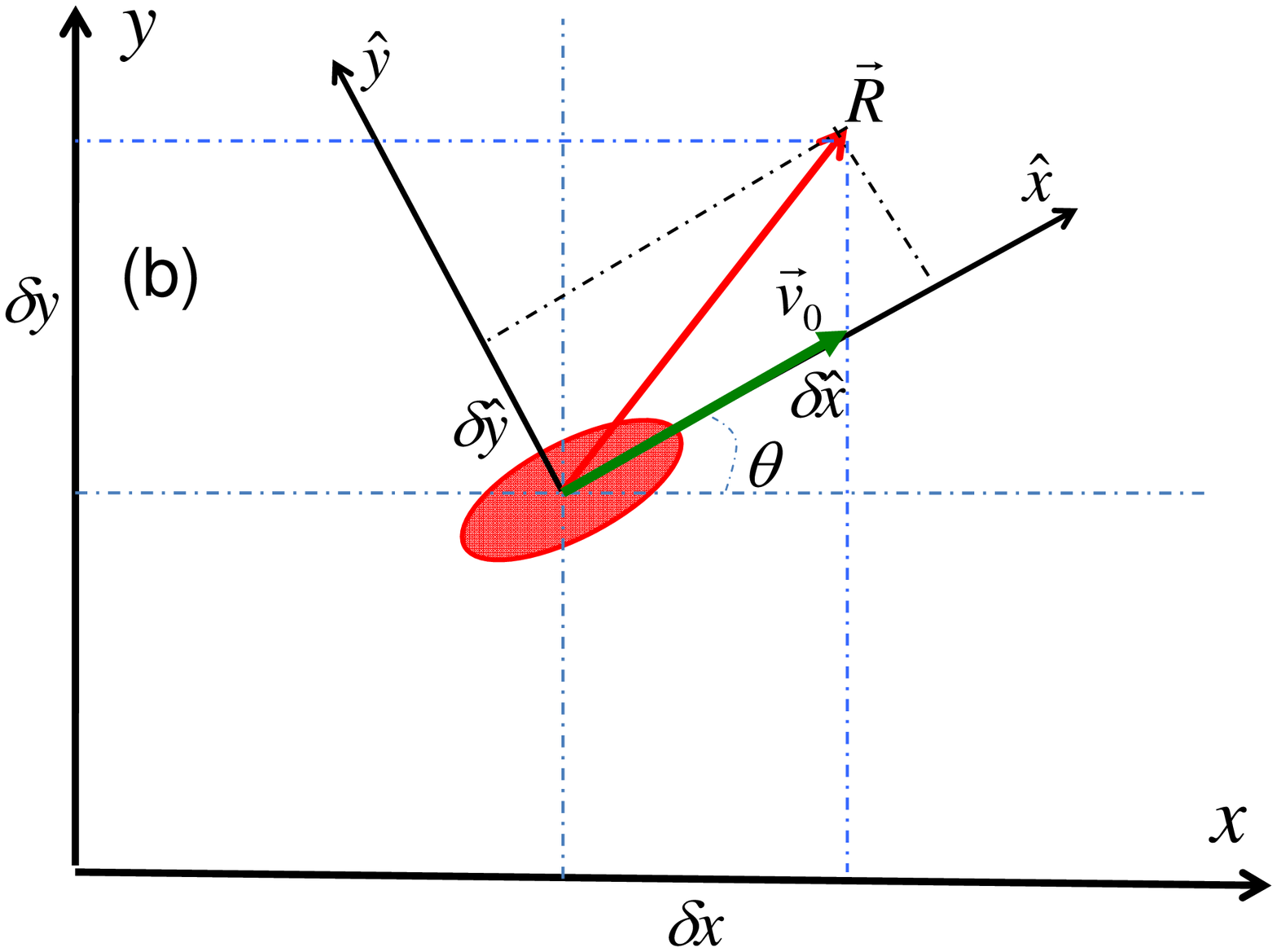}
  \vspace{-1.3cm}
\caption{Schematic diagram illustrating the system. (a) The active particle with the self-propelled velocity moving in the ratchet potential and the equipotentials of the potential described by Eq.(\ref{potential}). (b)Representation of an
ellipsoid in the $x$-$y$ lab frame and the $\hat{x}$-$\hat{y}$ body frame.  The angle between two frames is $\theta$. The
displacement $\vec{R}$ can be decomposed as ($\delta\hat{x}$, $\delta\hat{y}$) or ($\delta x$, $\delta y$).}\label{fig2}
\end{figure}


\begin{equation}\label{eq1}
  \frac{1}{\Gamma_1}\frac{\partial \hat{x}}{\partial t}= F_x\cos \theta(t)+F_y\sin\theta(t)+\hat{\xi}_1(t)+\frac{v_0}{\Gamma_1}
\end{equation}

\begin{equation}\label{eq2}
\frac{1}{\Gamma_2}\frac{\partial \hat{y}}{\partial t}=F_y\cos\theta(t)-F_x\sin\theta(t)+\hat{\xi}_2(t),
\end{equation}

\begin{equation}\label{eq3}
  \frac{1}{\Gamma_3}\frac{\partial \theta(t)}{\partial t}=\tau+\hat{\xi}_3(t),
\end{equation}
where $v_0$ is the self-propelled velocity in the body frame, which is taken along the long axis of the ellipsoid particle.
$\Gamma_1$ and $\Gamma_2$ are the mobilities along its long and short axis, respectively. $\Gamma_3$ is the rotational mobility and $\tau$ is the torque acting on the body due to it orientation relative to the direction of the potential. $F_x$ and $F_y$ are the forces along $x$ and $y$ direction of the lab frame.  The noise $\hat{\xi}_{i}(t)$ has mean zero and satisfies
\begin{equation}\label{}
  \langle \hat{\xi}_i(t)\hat{\xi}_j(t^{'})\rangle=\frac{2k_BT}{\Gamma_i}\delta_{i,j}\delta(t-t^{'}), i,j=1,2,3,
\end{equation}
where $T$ is the temperature and $k_B$ is the Boltzmann constant.

\indent We now obtain these equations in the lab frame based on the method described in Ref.\cite{Grima}. By means of a straight forward rotation of coordinates, the displacement in the two frames are related by the following equations,
\begin{equation}\label{}
  \delta x=\cos\theta \delta\hat{x}-\sin\theta\delta\hat{y},
\end{equation}
\begin{equation}\label{}
  \delta y=\sin\theta \delta\hat{x}+\cos\theta\delta\hat{y}.
\end{equation}
\indent After somewhat manipulations, Eqs. (\ref{eq1},\ref{eq2},\ref{eq3}) in the body frame can be replaced by the following equations in the lab frame \cite{Grima,Han}
\begin{equation}\label{eq11}
\frac{\partial x}{\partial t}=v_0\cos\theta(t)+F_x\bigg[\bar{\Gamma}+\Delta\Gamma\cos 2\theta(t) \bigg]+\Delta\Gamma F_y\sin 2\theta(t) +\xi_{1}(t),
\end{equation}

\begin{equation}\label{eq12}
  \frac{\partial y}{\partial t}=v_0\sin\theta(t)+F_y\bigg[\bar{\Gamma}-\Delta\Gamma\cos 2\theta(t) \bigg]+\Delta\Gamma F_x\sin 2\theta(t) +\xi_{2}(t),
\end{equation}

\begin{equation}\label{eq13}
 \frac{\partial \theta(t)}{\partial t}=\Gamma_3\tau+\xi_{3}(t),
\end{equation}
where the quantities $\bar{\Gamma}=\frac{1}{2}(\Gamma_1+\Gamma_2)$ and $\Delta\Gamma=\frac{1}{2}(\Gamma_1-\Gamma_2)$ are the average and difference mobilities of the body, respectively. The parameter $\Delta\Gamma$ determines the asymmetry of the body, the particle is a perfect sphere for $\Delta \Gamma=0$ and a very needlelike ellipsoid for $\Delta \Gamma\rightarrow\bar{\Gamma}$. The noise $\xi_{i}(t)$ has mean zero and the following relations \cite{Grima,Han}
\begin{equation}\label{}
   \langle \xi_3(t)\xi_3(t^{'})\rangle=2D_\theta \delta (t-t^{'}),
\end{equation}

\begin{equation}\label{}
 \langle \xi_i(t)\xi_j(t^{'})\rangle_{\theta(t)}^{\xi_1,\xi_2}=2k_{B}T\Gamma_{ij}\delta (t-t^{'}),
\end{equation}
and
\begin{equation}\label{}
  \Gamma_{ij}=\bar{\Gamma}\delta_{ij}+\Delta\Gamma \left(
                                                     \begin{array}{cc}
                                                       \cos 2\theta & \sin 2\theta \\
                                                       \sin 2\theta & -\cos 2\theta \\
                                                     \end{array}
                                                   \right),
\end{equation}
where $D_\theta=k_{B}T\Gamma_3$ is the rotational diffusion rate,
which describes the nonequilibrium angular fluctuation. The
statistical averages have superscripts to indicate over which noise
is the average taken and subscripts to denote quantities which are
kept fixed.

\indent For the asymmetric potential, we consider the following potential \cite{potential} (shown in Fig. 1(a)),
\begin{equation}\label{potential}
  U(x,y)=\frac{U_0}{2}y^2[\cos(x+\Delta\ln\cosh y)+1.1]+fx,
\end{equation}
where $U_0$ is the height of the potential and $f$ is the load along the $x$ direction. $\Delta $ is the asymmetric parameter of the potential and the potential is symmetric at $\Delta=0.0$. The equipotentials are now symmetry broken and look like a herringbone pattern for $\Delta\neq 0$.

\indent In this paper, we focus on the direction transport of active asymmetrical particles. The behavior of the quantities of interest can be corroborated
by Brownian dynamic simulations performed by integration of the Langevin equations (\ref{eq11},\ref{eq12},\ref{eq13}) using the second-order stochastic Runge-Kutta algorithm. Because the particle along the $y$-direction is confined, we only calculate the $x$-direction average velocity based on Eqs. (\ref{eq11},\ref{eq12},\ref{eq13}),
\begin{equation}\label{V}
            v_{\theta_0}=\lim_{t\rightarrow\infty}\frac{\langle x(t)\rangle^{\xi_1,\xi_2}_{\theta_0}}{t},
            \end{equation}
where $\theta_0$ is initial angle of the trajectory. The full average velocity after a second average over all $\theta_0$ is
\begin{equation}\label{}
  v=\frac{1}{2\pi}\int_0^{2\pi}d\theta_0 v_{\theta_0}.
\end{equation}
For the convenience of discussion, we define the scaled average velocity $v_s=v/v_0$ through the paper. In our simulations, the integration step time $\Delta t$ was chosen to be smaller than $10^{-4}$ and the total integration time was more than $3\times 10^5$ and the transient effects were estimated and subtracted. The stochastic averages reported above were obtained as ensemble averages over $3 \times 10^{4}$ trajectories with random initial conditions.

\section{Results and Discussion}
\indent Based on the numerical simulations, we mainly calculate the average velocity for the two cases: zero load and finite load.  For the zero load case, we focus on the rectification effects and how the parameters can affect rectification.  For the finite load case, we present two particle separation methods: shape separation and self-propelled velocity separation. In the simulations, unless otherwise noted, we set $k_BT=1.0$, $\tau=0.0$, $\bar{\Gamma}=1.0$, and $U_0=1.0$ throughout the paper.

\subsection{Zero load and rectification}

\begin{figure}
  \centering
  \includegraphics[width=8cm]{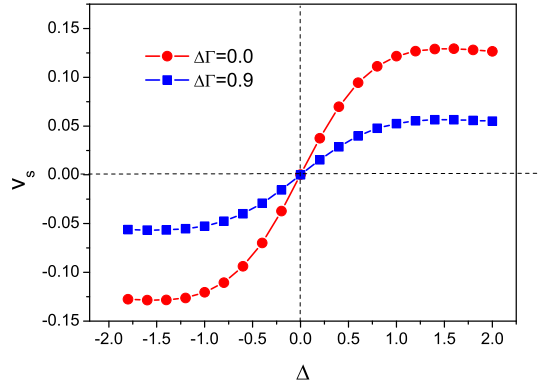}\\
  \caption{Scaled average velocity $v_s$ as a function of the asymmetrical parameter $\Delta$ of the potential for different values of $\Delta \Gamma$ at $v_0=2.0$ and $D_{\theta}=0.1$.}\label{fig2}
\end{figure}

\indent We first consider the zero load case ($f=0.0$). The scaled average velocity $v_s$ as a function of asymmetrical parameter $\Delta$ of the potential is reported in Fig. 2.  It is found that $v_s$ is positive for $\Delta>0$, zero at $\Delta=0$, and negative  for $\Delta<0$.  A qualitative explanation of this behavior can be given by the following argument. For $\Delta=0$ (symmetric case ) the probabilities of crossing right and left barriers are the same and then there is a null net particles flow. For $\Delta>0$, the left side from the minima of the potential is steeper, it is easier for particles to move toward the gentler slope side than toward the steeper side, so the average velocity is positive. Therefore, the asymmetry of the potential will determine the direction of the transport and no directed transport occurs in a symmetric potential.

\begin{figure}
  \centering
  \includegraphics[width=8cm]{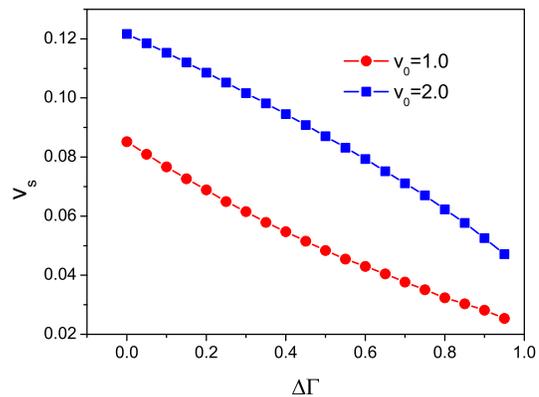}\\
  \caption{Scaled average velocity $v_s$ as a function of the asymmetrical parameter $\Delta \Gamma$ of the particle for different values of $v_0$ at $\Delta=1.0$, $f=0.0$, and $D_{\theta}=0.1$.}\label{fig3}
\end{figure}

\indent The dependence of the scaled average velocity $v_s$ on the asymmetrical parameter $\Delta\Gamma$ of the particle is presented in Fig. 3 at $\Delta=1.0$. We find that $v_s$ decreases monotonically with the increase of the asymmetrical parameter $\Delta\Gamma$. In order to give the explanation of the phenomenon, we present the translational diffusion coefficient in the $x$ direction \cite{Grima} $D_{xx}=D_0+\frac{\Delta\Gamma^{2}(F_{x}^2+F_{y}^2)}{8D_{\theta}}$, where $D_0=k_{B}T \bar{\Gamma}$. As we know, the increase of $D_{xx}$ enhances the ratchet effect when $D_{xx}<U_0$ and reduces the ratchet effect when $D_{xx}>U_0$. When $D_{xx}\simeq U_0$, the optimized ratchet effect occurs. In our system, $D_0=1.0$ and $U_0=1.0$, the ratchet effect is optimized when $\Delta\Gamma=0$ ($D_{xx}=U_0$).
 As $\Delta\Gamma$ increases from zero, $D_{xx}>U_0$, the ratchet effect is gradually destroyed and $v_s$ decreases monotonically.
Therefore, the perfect sphere particle can facilitate the rectification, while the needlelike ellipsoid particle destroys the directed transport.

\indent From Fig. 3, we can also find that the curve is convex for large value $v_0$ and concave for small value $v_0$. This phenomenon can be easily explained by introducing the two factors: (A) the increase of $v_0$ (from $0$ to $2$) enhances the transport (shown in Fig. 4) and (B) the increase of $\Delta\Gamma$ reduces the transport. For small value $v_0$ ($v_0=1.0$), on increasing $\Delta\Gamma$, factor B firstly dominates the transport, $v_s$ reduces quickly, and finally the ratchet effect gradually disappears (very small value $v_s$), $v_s$ reduces slowly,  so the curve is concave.  For $v_0=2.0$, on increasing $\Delta\Gamma$, factor A firstly determines the transport, $v_s$ reduces slowly, and finally factor B also becomes important, $v_s$ reduces quickly, so the curve is convex.
\begin{figure}
  \centering
  \includegraphics[width=8cm]{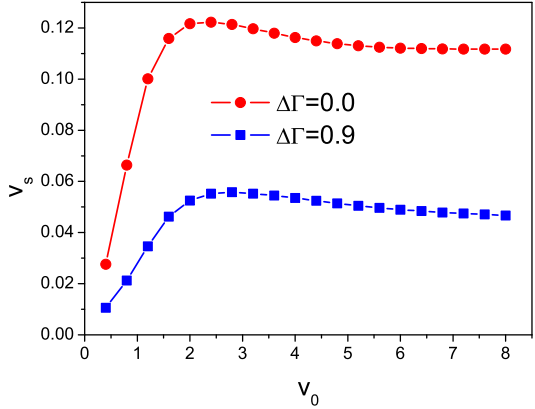}
  \caption{Scaled average velocity $v_s$ as a function of the self-propelled velocity $v_0$  for different values of $\Delta\Gamma$ at $\Delta=1.0$,$f=0.0$, and $D_{\theta}=0.1$.}\label{fig4}
\end{figure}

\indent Figure 4 shows the scaled average velocity $v_s$ as a function of the self-propelled velocity $v_0$. The term $v_0\cos\theta(t)$ in Eq. (\ref{eq11})  can be seen as the external driving force.  When $v_0\rightarrow 0$, the external driving force can be negligible, so the average velocity will tend to zero. For very large values of $v_0$, the effect of the asymmetry of the potential reduces, thus $v_s$ becomes small. Therefore, there exists an optimal value of $v_0$ at which $v_s$ takes its maximal value.  So the optimal self-propelled velocity can facilitate the rectification of particles.

\begin{figure}
  \centering
  \includegraphics[width=8cm]{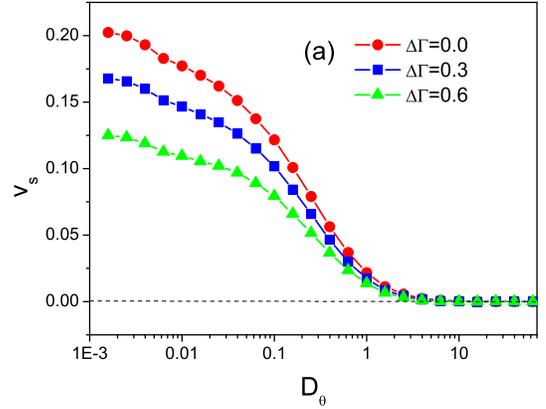}
   \includegraphics[width=8cm]{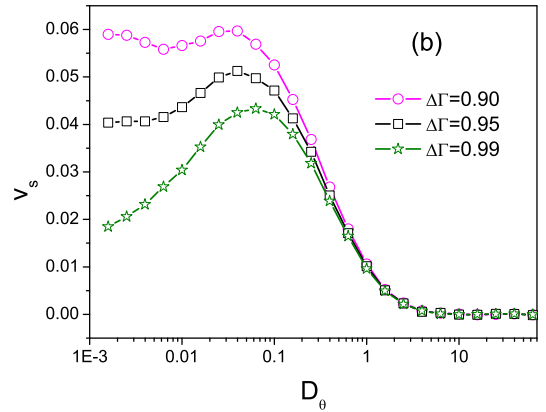}
  \caption{Scaled average velocity $v_s$ as a function of the rotational diffusion rate $D_\theta$ for different values of $\Delta\Gamma$. (a)ellipsoidal particles ($\Delta\Gamma=0, 0.3, 0.6$). (b)needlelike particles ($\Delta\Gamma=0.9, 0.95, 0.99$). The other parameters are $\Delta=1.0$, $f=0.0$, and $v_0=2.0$.}\label{fig5}
\end{figure}
\indent Results for $v_s$ as a function of the rotational diffusion rate $D_{\theta}$ are presented in Fig. 5 for different values of $\Delta\Gamma$.
For the case of ellipsoidal particles ( $\Delta\Gamma=0, 0.3, 0.6$) shown in Fig. 5(a),  the scaled average velocity $v_s$ decreases with the increase of $D_{\theta}$. The term $v_0\cos\theta(t) (\propto v_0\cos(D_{\theta} t)) $ in Eq. (\ref{eq11}) can be seen as the external driving force along $x$-direction.
In the adiabatic limit $D_\theta\rightarrow 0$, the external force can be expressed by two opposite static forces $v_0$ and $-v_0$, yielding the mean velocity $V=\frac{1}{2}[v(v_0)+v(-v_0)]$, which is similar to the adiabatic case in the forced thermal ratchet \cite{rmp}.  As $D_\theta$ increases, the scaled average velocity $v_s$ decreases. When $D_\theta\rightarrow\infty$, the self-propelled angle changes very fast, particles are trapped in the valley of the potential, so $v_s$ tends to zero, which is similar to the high frequency driving case in the forced thermal ratchet \cite{ai}.

 \indent However, for the case of needlelike particles ($\Delta\Gamma=0.9, 0.95, 0.99$) shown in Fig 5 (b), the anisotropic diffusion dominates the transport and there exists an optimal value of $\Delta\Gamma$ at which $v_s$ takes its maximal value. This is due to the mutual interplay between the anisotropic diffusion and the rotational diffusion rate. There are two time periods in the system: the anisotropic diffusion time $\tau_D=\frac{4\pi^2}{D_{xx}}$ for crossing one period of the potential along $x$-direction and the period $\tau_{\theta}=\frac{2}{D_{\theta}}$ for direction randomly varying in time. For the case of needlelike particles, the anisotropic diffusion time becomes very important. When these two time periods cooperate with each other, the optimized rectification will occur.

\begin{figure}
  \centering
  \includegraphics[width=8cm]{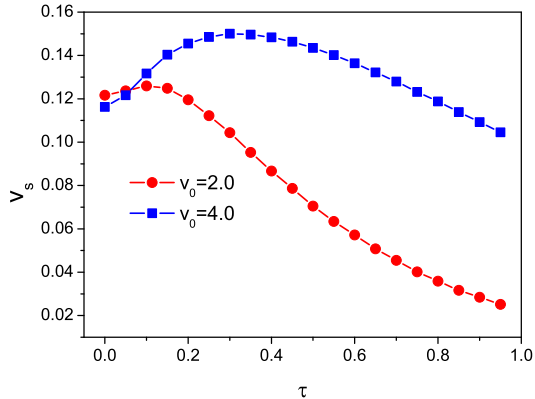}\\
  \caption{Scaled average velocity $v_s$ as a function of the torque $\tau$ acting on the body  for different values of $v_0$ at $\Delta=1.0$,$f=0.0$, $D_{\theta}=0.1$, and $\Delta\Gamma=0.0$.}\label{fig6}
\end{figure}

\indent Figure 6 describes the dependence of the scaled average velocity $v_s$ on the torque $\tau$ for different values of $v_0$.  From Eq. (\ref{eq13}), we can see that the self-propelled angle $\theta(t)$ is determined by the torque and the random noise.  When $\tau\rightarrow 0$, the rotational diffusion rate determines the angle $\theta(t)$. On increasing $\tau$, both the torque and the random noise play the important roles and work together in the transport, which induces the maximal average velocity.  However, when $\tau\rightarrow \infty$, the self-propelled angle changes very fast, particles are trapped in the valley of the potential and $v_s$ goes to zero. Therefore, large torque would suppress strongly the ratchet effect.

\subsection{Finite load and particle separation}

\indent Since transport behaviors in the present system
strongly depend on the asymmetric parameter $\Delta \Gamma$ of the particle and the self-propelled velocity $v_0$ , it is possible
to realize particle separation. We will present two particle separation mechanisms which induce the motion of particles of different $\Delta\Gamma$ or $v_0$ in opposite directions by introducing an external load $f$ on the $x$-direction.
\begin{figure}
  \centering
  \includegraphics[width=8cm]{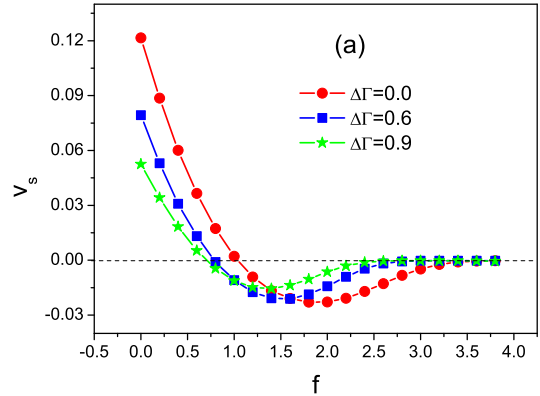}
   \includegraphics[width=8cm]{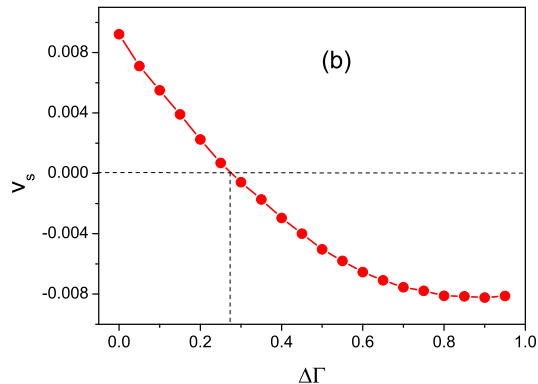}
  \caption{(a) Scaled average velocity $v_s$ as a function of the load $f$ for different values of $\Delta\Gamma$. (b)Scaled average velocity $v_s$ as a function of the parameter $\Delta\Gamma$ at $f=0.9$. The other parameters are $\Delta=1.0$ and $v_0=2.0$, and $D_{\theta}=0.1$.}\label{fig3}
\end{figure}

\indent Figure 7 (a) shows the scaled average velocity as a function of the load $f$ for different values of $\Delta\Gamma$.  For small values of $f$, the positive ratchet effect dominates the transport and the average velocity is positive. On increasing the load $f$, the load dominates the transport, the average velocity crosses zero and subsequently reverses its direction.  For very large load, particles are blocked and $v_s$ tends to zero exponentially.
From Fig. 1(a), we can find that if the load $f$ is negative (positive force along $x$-direction), the particle will move forward without problem. If the load $f$ is positive (negative force along $x$-direction), the particle will move backward and may get into a spine of the herringbone.  Thus, the particle has to go against the force in order to climb back up and get again on the backbone. Moreover, one can find that the stronger the force, the deeper the spine. Therefore, the particle is blocked in the spines for very large values of $f$. Note that this blocked phenomenon had been explained detailedly in Ref. \cite{potential}. There exists a valley in the curve $v_s-f$ at which the average velocity takes its negative maximal value.  The position of the valley varies with $\Delta\Gamma$.  For a given load ($f=0.9$) shown in Fig. 7 (b), the asymmetric parameter $\Delta\Gamma$ of the particle can determine the direction of the transport. Particles larger than the threshold asymmetric parameter $\Delta\Gamma_c$  move to the left, whereas particles smaller than that move to the right. Therefore, one can separate particles of different values of $\Delta\Gamma$ and make them move in opposite directions.
\begin{figure}
  \centering
  \includegraphics[width=8cm]{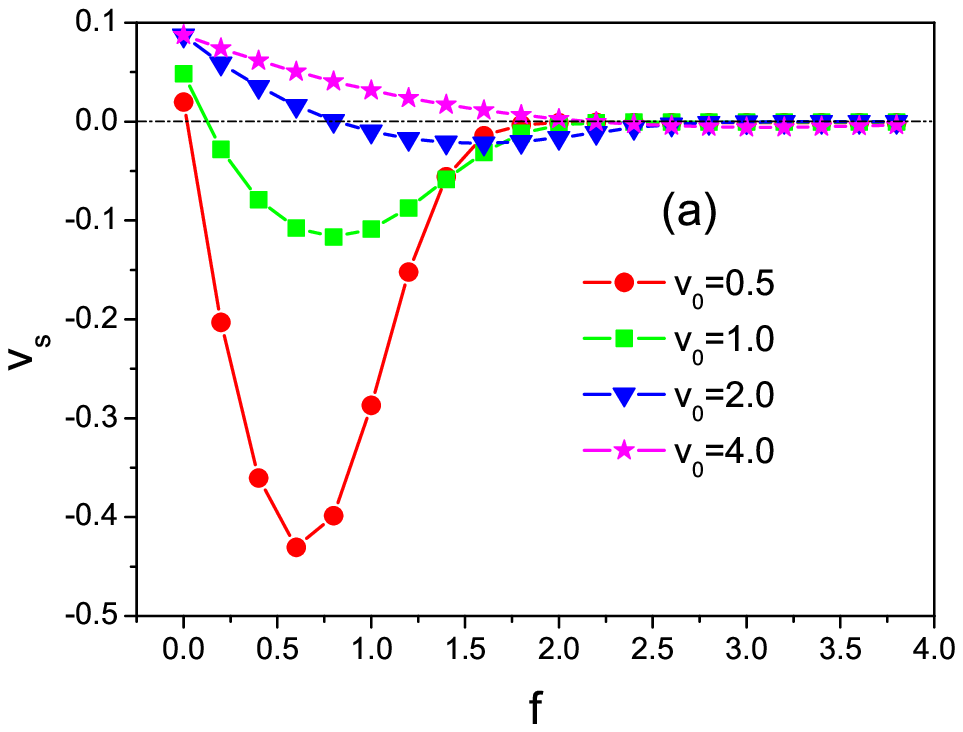}
   \includegraphics[width=8cm]{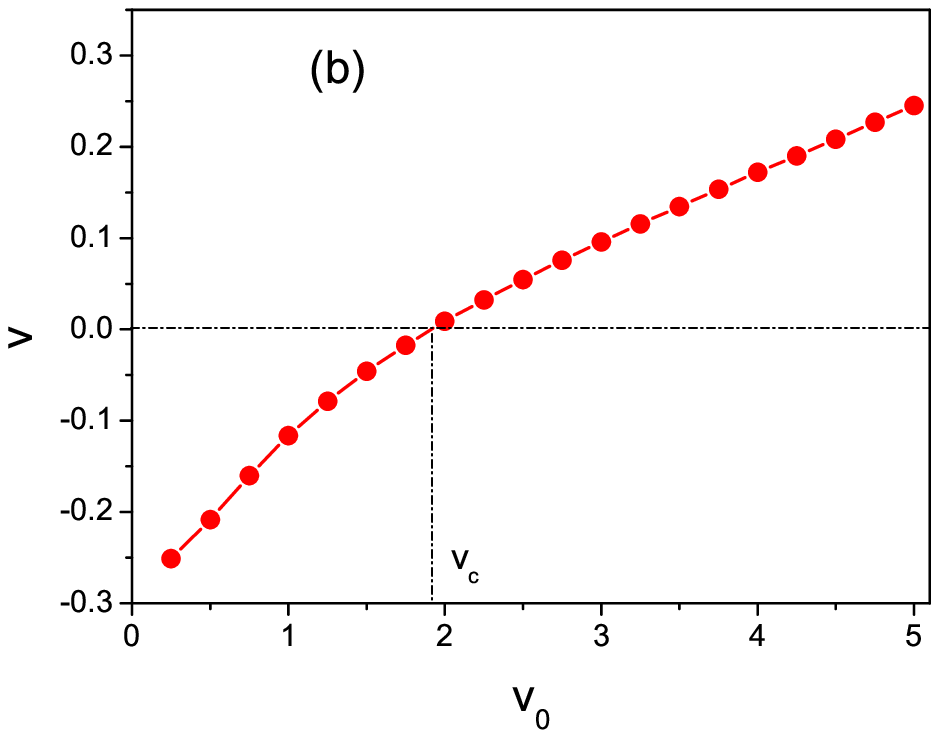}
  \caption{(a) Scaled average velocity $v_s$ as a function of the load $f$ for different values of $v_0$. (b) Average velocity $v$ as a function of $v_0$ at $f=0.75$. The other parameters are $\Delta=1.0$ and $\Delta\Gamma=0.5$, and $D_{\theta}=0.1$.}\label{fig3}
\end{figure}

\indent Figure 8 (a) shows the scaled average velocity as a function of the load $f$ for different values of the  self-propelled velocity $v_0$.
There exist two driving factors in the system: (1) the self-propelled velocity $v_0$ which induces the positive current; (2) the load $f$ which causes the negative current. For small values of $v_0$ (e. g. $v_0=0.5$), the load $f$ dominates the transport and the self-propelled velocity $v_0$ can be neglected.  When $f\rightarrow 0$, the average velocity goes to zero. When increasing $f$, the amplitude of the negative velocity increases. However, for very large values of $f$, the particle is blocked and the average velocity tends to zero. Therefore, there exists a valley in the curve $v_s-f$ at which the average velocity takes its negative maximal value.  However, on increasing $v_0$, the self-propelled velocity driving factor becomes more important and the positive current gradually dominates the average velocity. Therefore, the valley in the curve gradually disappears.  Therefore, the self-propelled velocity $v_0$ strongly affects the transport and even determines its direction. For a given value of $f$ (e. g. 0.75) shown in Fig. 8 (b), particles  with larger than the threshold velocity $v_c$ move to the right, whereas particles smaller than that move to the left. Therefore, one can separate particles of different values of $v_0$ and make them move in opposite directions.

\indent Note that there are some other methods which can separate particles for our system. Particles can be separated if their drift speeds are different, no need to move in the opposite directions. Particles with different shapes can be separated by many other means rather than the ratchet method.  However, our separation methods (which make particles move in opposite directions) are more effective and feasible.

\section {Concluding remarks}
\indent In this work we have studied the transport of active ellipsoidal particles in a two-dimensional asymmetric potential by numerical simulations. It is found that the self-propelled velocity can break thermodynamical equilibrium and induce directed transport. The direction of the transport is determined by the asymmetry of the potential. The shape of particles can strongly affect the rectified transport, the perfect sphere particle can facilitate the rectification, while the needlelike particle destroys the directed transport. There exist optimized values of the parameters (the self-propelled velocity, the torque acting on the body) at which the average velocity takes its maximal value. For the ellipsoidal particle with not large asymmetric parameter, the average velocity decreases with increasing the rotational diffusion rate, while for the needlelike particle (very large $\Delta\Gamma$ ), the average velocity is a peaked function  of the rotational diffusion rate. In addition, by introducing a finite load on particles, we have presented two particle separation ways (1) shape separation: particles larger than the critical asymmetric parameter $\Delta\Gamma_c$  move to the left, whereas particles smaller than that move to the right; (2)self-propelled velocity separation: particles with larger than the threshold velocity $v_c$ move to the right, whereas particles smaller than that move to the left. Therefore, one can separate particles of different shapes (or different self-propelled velocities) and make them move in opposite directions. The results we have
presented have a wide application in many systems, such as spontaneously moving oil droplets, motor proteins, bacterial swimmers, and motile cells.

\indent  This work was supported in part by the National Natural Science Foundation
of China (Grant No. 11175067), the PCSIRT (Grant No. IRT1243), the Natural
Science Foundation of Guangdong Province, China (Grant No. S2011010003323), the Scientific Research Foundation of Graduate School of South China Normal University.

\end{document}